\documentclass[aps,floats,floatfix,showpacs,amssymb,prd,twocolumn,superscriptaddress,nofootinbib,longbibliography,reprint]{revtex4-1}
\usepackage{amssymb,amsmath,verbatim,mathtools,needspace,enumitem,etoolbox,graphicx,physics,microtype,afterpage,xspace,tabularx,lmodern,multirow}
\usepackage{gensymb}
\usepackage[normalem]{ulem}
\usepackage[dvipsnames, usenames]{xcolor}
\definecolor{linkcolor}{rgb}{0.0,0.3,0.5}
\usepackage[unicode, colorlinks=true, linkcolor=linkcolor, citecolor=linkcolor, filecolor=linkcolor, urlcolor=linkcolor, linktocpage, breaklinks]{hyperref}
\usepackage[all]{hypcap}
\usepackage[T1]{fontenc}
\usepackage[utf8]{inputenc}
\usepackage[usenames,dvipsnames]{xcolor}

\hypersetup{colorlinks=true,citecolor=romared,linkcolor=romared,urlcolor=romared}
\definecolor{romared}{RGB}{142,0,28}
\usepackage{aas_macros}
\usepackage{makecell}
\usepackage{soul}
\usepackage{booktabs}
\newcommand{\OU}{\affiliation{Department of Physics, Oakland University, Rochester, Michigan 48309, USA}}

\usepackage{orcidlink}

\begin{document}
\title{On the orbital eccentricities of primordial black hole binaries inside and outside of dark matter halos} 

\begin{abstract}
Primordial black hole (PBH) binaries in the stellar mass range may still contribute a fraction of the detectable compact object binaries by LIGO and future GW observatories. PBH binaries at formation typically have very high eccentricities. In this paper, we study the eccentricity of stellar mass range PBH binaries from all formation channels and account for all evolutionary pathways. We simulate large samples of PBH binaries, tracking their full orbital evolution up to their merger or to the present day. For those that merge, we compute their GW strain, detectability, and eccentricity distributions for LISA, DECIGO, ET, CE, and aLIGO.  We find that PBH binaries that evolve in isolation completely circularize by the time their GWs enter any GW band except for LISA's, where residual eccentricities of order  $O(10^{-2})$ can exist. Binaries that become part of dark matter halos can have multiple binary-single interactions with other PBHs, especially if they reside in the more dense environments among them and can have higher eccentricities even at their late inspiral phase, probed by the GW observatories. Considering the current limits on the abundance of stellar mass range PBHs, we predict that LISA and DECIGO together would be able to probe $O(10^2)$ such binaries with $e>0.01$. If these future GW observatories in space can exclude such eccentric binaries, then limits on the PBH abundance can be improved by an order of magnitude. 

\end{abstract}

\author{Muhsin Aljaf\,\orcidlink{0000-0002-1251-4928}}
\email{muhsinaljaf@oakland.edu}
\OU

\author{Ilias Cholis\,\orcidlink{0000-0002-3805-6478}}
\email{cholis@oakland.edu}
\OU

\date{\today}
\maketitle

\section{Introduction}\label{intro}
Primordial black holes (PBHs) are compact objects that could have been produced from the collapse of early universe large density perturbations~\cite{1967SvA....10..602Z, Hawking:1971ei, Carr:1974nx}. Depending on their mass distribution, PBHs may be able to explain some fraction and up to all of the detected dark matter \cite{Carr:2016drx, Carr:2017jsz, Carr:2020gox}. Furthermore, for certain assumptions on their mass, some of these PBHs could be in binaries and merge, giving detectable gravitational wave (GW) signals \cite{Sasaki:2016jop, Ali-Haimoud:2017rtz}. 
With the first GW event detected by LIGO in 2015~\cite{LIGOScientific:2016aoc}, PBHs in the solar mass range, $\mathcal{O}(10)\ M_\odot$, have been proposed as one of the mechanisms by which merging black hole binaries may come to form \cite{Bird:2016dcv,Sasaki:2016jop,Sasaki:2018dmp}. 
Since then, GW observations have put constraints on the PBH abundance in the universe, limiting it to a fraction, $f_{\rm PBH} \equiv \Omega_{\rm PBH}/\Omega_{\rm DM}\lesssim 10^{-2}$ within this solar mass range~\cite{Sasaki:2016jop, Ali-Haimoud:2017rtz, Kavanagh:2018ggo, Hutsi:2020sol, Andres-Carcasona:2024wqk, Bouhaddouti:2025ltb, Bouhaddouti:2026jgc}, with other limits for $O(10)$-$O(100)$ $M_{\odot}$ PBHs described in Refs.~\cite{Macho:2000nvd, EROS-2:2006ryy, Ricotti:2007au, 2014ApJ...790..159M, Chen:2016pud, Ali-Haimoud:2016mbv, Brandt:2016aco, Horowitz:2016lib, Poulin:2017bwe, Zumalacarregui:2017qqd, Serpico:2020ehh, Green:2020jor, Mroz:2024wag, Mroz:2024mse}.

Most PBH binaries were formed in the early universe when two nearby PBHs decoupled from the Hubble flow around the matter-radiation equality era \cite{Nakamura:1997sm, Ioka:1998nz, Sasaki:2016jop, Raidal:2018bbj}. 
At that stage, typically PBH binaries got their angular momentum from the torques of neighboring single PBHs, making their orbits highly eccentric at formation.
Such binaries evolved solely via GW emission. 
However, as halos with significant mass started to form around $z=12$ \cite{Press:1973iz}, two different evolutionary paths emerged. One has binaries that remained outside dark matter halos, continuing their evolution purely via GW emission, with their orbits getting circularized over very long time scales. 
We call these unperturbed binaries \cite{Sasaki:2016jop, Ali-Haimoud:2017rtz}. 
The other evolutionary path contains binaries that fell in halos and underwent stochastically binary-single interactions with neighboring PBHs, which could have hardened or softened their orbits and caused their eccentricity to grow or decay, making their merger times shorter or longer depending on the properties of their host halos~\cite{Aljaf:2024fru, Aljaf:2025dta}. We refer to these interactions as binary-single interactions. 

The second mechanism by which PBH binaries can form at any redshift is when two unbound single PBHs in dark matter halos lose energy via GW emission during a close encounter, forming a bound binary as shown in  Ref.~\cite{Bird:2016dcv} (for an updated calculation see Ref.~\cite{Aljaf:2024fru}). The formation channel from such a mechanism is typically referred to as direct captures. Through that mechanism, a small fraction of the formed binaries come from near ``head-on collisions'' of PBHs. Thus, such binaries can be characterized by high orbital eccentricities, and can lead to rapid mergers \cite{Cholis:2016kqi}.

There are three main properties of a binary that are related to how it formed and evolved: its component masses, its spins, and its orbital eccentricity. 
In particular, the eccentricity when the GW signal enters a detector is associated with the binary’s formation and evolution history~\cite{Mandel:2018hfr}. 
Early-formed binaries that evolved in isolation are expected to have very small eccentricity by the time they reach ground-based detectors as is the current LIGO-Virgo-KAGRA(LVK) \cite{LIGOScientific:2014pky}, the future Einstein Telescope (ET) \cite{ET:2025xjr}, and the future Cosmic Explorer (CE) \cite{Evans:2021gyd} which (in the case of ET) will reach detector-frame gravitational frequencies of $f_{\rm det}\gtrsim1$ Hz. 
Early-formed PBH binaries are expected to have non-detectable eccentricities in ground-based detector frequencies as GW emission circularizes the orbits on a timescale much smaller than the lifetime of the universe~\cite{PhysRev.131.435, Franciolini:2021xbq}. 
In contrast, in dense environments such as massive star clusters and the inner parts of dark matter halos, PBH binaries undergoing binary-single interactions can have their merger timescale significantly reduced.  Direct PBH captures can create binaries with very small merger timescales \cite{Bird:2016dcv}, and thus highly eccentric orbits \cite{Cholis:2016kqi}. 
Some of the PBH binaries in dark matter halos may not have enough time for the binary's orbit to fully circularize before entering GW ground-based detector frequencies. 
A non-negligible amount of eccentricity can therefore be taken as evidence of environmental interactions ~\cite{Samsing:2017xmd, Samsing:2017oij, Lower:2018, Wang:2020jsx, Romero-Shaw:2021, Kritos:2020wcl, Samsing:2024syt, Hendriks:2024zbu, Fabj:2024jns, Fabj:2025vza}. 
At lower frequencies ($f_{\rm det}<10^{-2}$ Hz), space-based observatories like the Laser Interferometer Space Antenna (LISA) \cite{LISA:2022yao} and the Deci-hertz Interferometer Gravitational wave Observatory (DECIGO) \cite{Kawamura:2020pcg} will be able to observe binaries at earlier stages of their inspiral, where eccentricity can still be non-negligible. This makes them especially interesting to study binary formation channels and evolutionary pathways~\cite{Breivik:2016, Gondan:2017wzd, Chen:2017, Franciolini:2021xbq, Wang:2024, Holst:2024ubt, Samsing:2024syt, Riotto:2024ayo, Fabj:2025vza}.

Despite several studies on PBH merger rates and mass distribution~\cite{Aljaf:2024fru,Aljaf:2025dta,Ali-Haimoud:2017rtz,Raidal:2018bbj,Sasaki:2018dmp,Bouhaddouti:2025ltb,Bouhaddouti:2026jgc}, a detailed study of the eccentricity distribution across multiple frequency bands is still missing for the environments PBH binaries are in.
In this work, we address this question. 
We track large samples of PBH binaries through their full orbital evolution, accounting for all formation channels and evolutionary pathways, computing their GW strain, detectability, and eccentricity distribution for LISA, DECIGO, ET, CE, and LIGO. We find that the PBH binaries that remain unperturbed fully circularize across all bands except LISA. However, binaries that undergo several binary-single interactions inside dense dark matter halo environments have higher eccentricities even in their late inspiral phases, and both LISA and DECIGO will be able to probe $O(10^2)$ binaries with an eccentricity larger than 0.01, a few among which will have an eccentricity larger than 0.1.
These numbers of PBH binaries with some remaining eccentricity take into account the current limits on the abundance of PBHs from the existing GW observations by the LVK collaboration (see Ref.~\cite{Bouhaddouti:2026jgc} for a recent update). 
While binaries from direct captures start with non-negligible eccentricities, given those same limits on the PBH abundance, this class of binaries has only a minor contribution to the total number of PBH binaries with an eccentricity larger than $0.01$.

The paper is organized as follows: Section~\ref{sec:methodology} describes the orbital evolution model for all three cases, the computation of the characteristic strain, and the detectability criterion. Section~\ref{sec:Results} presents our results: eccentricity distributions and characteristic-strain tracks for each channel, together with the expected detection counts.  We conclude in Section~\ref{sec:Conclusions}.

\section{Methodology}\label{sec:methodology}
In this section, we discuss how we model the evolution and detectability of PBH binaries from their formation to their last stable orbit before their merger. 
Following the methodology of Ref.~\cite{Aljaf:2024fru, Aljaf:2025dta}, we classify PBH binaries into three evolutionary pathways: 
i) early binaries that formed just after the individual PBHs' formation and whose orbital properties evolved from that point on through GW emission alone, as throughout their remaining history they remained unperturbed, 
ii) perturbed binaries, i.e., early PBH binaries that at some point in time became part of a dark matter halo, where they underwent further binary–single interactions that affected the evolution of their orbital properties, and 
iii) late PBH binaries that formed by direct GW captures inside dark matter halos, oftentimes with very high eccentricities. 
In our simulations, we first study each evolutionary path, ``channel'' separately by tracking large numbers of PBH binaries and evolving the semi-major axis and eccentricity of each of these. 
We then compute the GW signal from each binary and check if it would be detectable by comparing its characteristic strain against the noise curves of current and future GW observatories and calculating its signal-to-noise ratio (SNR).
Finally, we appropriately combine the three channels to evaluate how many PBH binaries with a non-zero eccentricity during their inspiral can be observed. 

\subsection{Orbital Evolution of PBH Binaries}\label{sec:PBH orbital evolution}
The orbital evolution of PBH binaries is determined by the interplay between GW emission and the interactions with the environment where the binary is located. 
The evolution of the semi-major axis $a$ and eccentricity $e$ of PBH binaries is given by \cite{1987Binney,Aljaf:2025dta}\footnote{Some binaries can soften as well, but those  are not part of the merging population.},
\begin{eqnarray}
\frac{da}{dt} &=&-\frac{G \, H \,\rho_{\rm env}(r, t)}{v_{\rm disp}^{\rm env}(r, t)}  a^2  + \frac{da}{dt}\bigg|_{\rm GW}, \label{eq:evol_a_hard} \\
\frac{de}{dt}&=&+\frac{G \, H \,K(r,t)\,\rho_{\rm env}(r, t)}{v_{\rm disp}^{\rm env}(r, t)}  a + \frac{de}{dt}\bigg|_{\rm GW},
\label{eq:evol_e_hard}
\end{eqnarray}
where $c$ is the speed of light and $G$ is the universal gravitational constant. The first terms in both equations come from  environmental interactions of PBH binaries within dark matter halos through $\rho_{\textrm{env}}$ and $v^{\textrm{env}}{\textrm{disp}}$. The coefficients $H$ (not to be confused with the Hubble function $H(z)$) and $K$ characterize the efficiency of these interactions as discussed in \cite{Quinlan:1996vp, Sesana:2006xw}.
These contributions vanish outside halos, reducing the binary evolution to purely GW-driven, i.e., following the orbit-averaged Peters–Mathews Equations of Ref.~\cite{PhysRev.131.435}. We assume for simplicity that the members of the PBH binaries have masses $m{1}$ and $m_{2}$ and the surrounding single PBHs have all the same mass, which we take to be $30 M_{\odot}$.

For the unperturbed channel, we sample $5 \times 10^6$ PBH binaries with initial orbital parameters $(a_0,e_0)$ at $z=3400$ from the distributions of Refs.~\cite{Franciolini:2021xbq,Kavanagh:2018ggo} (see also~\cite{Sasaki:2016jop,Ali-Haimoud:2017rtz}). For these binaries we solve numerically Eqs.~\eqref{eq:evol_a_hard}–\eqref{eq:evol_e_hard} (ignoring the environmental terms), with an adaptive Runge–Kutta solver using \texttt{SciPy}'s \texttt{Solve\_ivp} routine\cite{2020SciPy-NMeth}, evolving each binary in the sample until it reaches its innermost stable circular orbit (ISCO), defined as $r_{\rm ISCO} = 3R_s = 6G(m_1+m_2)/c^2$.
The resulting evolution of $a(t)$ and $e(t)$ for the binaries is then used for the characteristic strain and SNR calculations described in sections~\ref{sec:Characteristic strain} and~\ref{subsec:SNR}.

For the binary–single interaction evolutionary channel, we simulate a much larger number of PBH binaries.
Based on earlier work in Ref.~\cite{Aljaf:2025dta}, we know for binary-single interactions inside halos  with present masses between $10^9,M_\odot \leq M \leq 10^{15},M_\odot$, that while they may have an effect on the total PBH merger rate, they are not common enough to increase very significantly the eccentricity of the PBH binaries \footnote{There are also binary-binary interactions that take place inside dark matter halos. In almost all cases, given the small fraction of hard PBH binaries in most of the volume of those halos, a hard-PBH binary will interact practically always with significantly softer PBH binaries. Such interactions can be treated as binary-single interactions from the perspective of the evolution of the hard binary.}.
For GW frequencies observable by ground-based and space detectors, we assume that the eccentricity distribution of PBH binaries inside halos that grow to a mass larger than $10^9 M_\odot$ is very similar to the eccentricity distribution of unperturbed binaries.
We focus our simulations on dark matter halos with present day masses spanning $10^4 M_\odot \leq M < 10^{9} M_\odot$.
For a given halo, we divide it into $N$ concentric spherical shells.
Each shell has its own time-dependent $\rho_{\rm env}$ and $v_{\rm disp}^{\rm env}$, which account for spatial variations within the halo across time, as the mass of each shell and the total halo mass grow with time.

For any given shell, we sample PBH binaries that enter the host halo and that specific shell gradually, following the halo mass growth (see \cite{Aljaf:2025dta} for more details). Before entering the halo, these PBH binaries evolve from their formation at $z= 3400$ to that point in time, purely through gravitational-wave emission, i.e., using only the GW terms in Eqs.~\eqref{eq:evol_a_hard}–\eqref{eq:evol_e_hard}.
Given that before $z = 12$, there is very little dark matter mass inside halos, we assume that binary-single interactions are negligible before $z=12$. To ensure a large sample of PBH binaries and properly probe their eccentricity distribution, for every combination of dark matter halo mass and shell, we track $5\times 10^{6}$ PBH binaries, all of which are formed at $z=3400$. For the smaller halo masses, that translates to the large number of simulated binaries becoming part of different dark matter halos. We properly re-weight our simulations by accounting for the mass and redshift-dependent halo mass function \cite{Press:1973iz} (see Ref.~\cite{Aljaf:2025dta} for further details).

In each simulation, for any shell, the halo environmental properties $\rho_{\rm env}$ and $v_{\rm disp}^{\rm env}$ are updated every $dt = 200,\rm{Myr}$.
Within each $dt$ step, the binary orbital parameters $(a,e)$ of each of the simulated binaries are integrated via the Euler method with a smaller time step $dt_{\rm local} = 0.1,\rm{Myr}$, advancing as $t_{\rm local} = t_{\rm local} + dt_{\rm local}$ until $t_{\rm local} = dt$.
When a PBH binary evolves inside the dark matter halo, to the point in time when the environmental terms of Eqs.~\eqref{eq:evol_a_hard}–\eqref{eq:evol_e_hard} become less important than the GW terms, we switch from that point on back to purely GW evolution.
That switch in our simulations happens when
the ratio of the magnitudes of the first term to the second term in Eq.~\eqref{eq:evol_a_hard} becomes $|\dot{a}|{\rm env}/|\dot{a}|{\rm GW} \leq 0.5$,
while also the equivalent condition for Eq.~\eqref{eq:evol_e_hard}, $|\dot{e}|{\rm env}/|\dot{e}|{\rm GW} \leq 0.5$ is satisfied. We define  that point in time as $t_{\rm GW}$ and record the $(a,e) = (a_{\rm GW}, e_{\rm GW})$.
At $t_{\rm GW}$, we pass the $(a_{\rm GW}, e_{\rm GW})$ to the Runge-Kutta solver in Python to evolve the binary forward using only the GW terms until it reaches its ISCO
\footnote{As the orbit shrinks, the dynamics accelerate ($\dot{a}\propto a^{-3}$). We switch to the adaptive Runge–Kutta solver method as we want to  ensure numerical precision during the late inspiral phase.}. The late stage evolution of $a(t)$ and $e(t)$ of the binaries is used for the evaluation of the characteristic strain and SNR described in sections~\ref{sec:Characteristic strain} and~\ref{subsec:SNR}.

\subsection{Characteristic strain}\label{sec:Characteristic strain}
For any binary in order to be observable, it needs to have entered the GW domination regime. For a given combination of $(a, e)$ along the binary's evolution, we can compute its GW signal at a detector. 
In the source frame, a binary emits radiation at discrete harmonics $n$, $f_{r}(n)$ of the Keplerian orbital frequency $f_{\textrm{orb}}$~\cite{PhysRev.131.435},
\begin{equation}
f_{r}(n) = n\, f_{\textrm{orb}}, \qquad n \ge 2,
\end{equation}
with,  
\begin{equation}
f_{\textrm{orb}}=\frac{1}{2\pi}\sqrt{\frac{G(m_1+m_2)}{a^3}}.
\end{equation}
The associated GW power radiated in the $n$-th harmonic is given by,
\begin{eqnarray}
P_n &=& P_{2} \; g(n, e) \nonumber \\
 &=& \frac{32}{5} \frac{G^4}{c^5} \frac{m_1^2 m_2^2 (m_1 + m_2)}{a^5} \; g(n, e),
\end{eqnarray}
where the function $g(n, e)$ quantifies the power enhancement factor of the $n$-th harmonic relative to the power emitted in the $n=2$ harmonic for a circular orbit and it is given by Eq.~20 of Ref.~\cite{PhysRev.131.435}, 
\begin{eqnarray}
g(n, e) &=& \frac{n^4}{32}\left\{\left[J_{n-2}(n e)-2 e J_{n-1}(n e)+\frac{2}{n} J_n(n e)\right.\right. \nonumber \\
&& \; \; \; \; \; \; \; \left.\; \; \; +2 e J_{n+1}(n e)-J_{n+2}(n e)\right]^2 \nonumber \\
&& \; \; \; \; \; \; \;\; \; \; \,+\left(1-e^2\right)\left[J_{n-2}(n e) -2 J_n(n e)\right. \nonumber \\
&& \; \; \; \; \; \; \; \; \; \; \left.\left.+J_{n+2}(n e)\right]^2+\frac{4}{3 n^2}\left[J_n(n e)\right]^2\right\},
\label{eq:g_ne}
\end{eqnarray}
where $J_n$ is the Bessel function of the first kind of the $n$-th order. 

It is useful to consider the peak harmonic, $n_{\text{peak}}(e)$, which represents the value of $n$ where $g(n, e)$ reaches its maximum. The associated frequency of GWs associated with the peak harmonic is then given by, 
\begin{equation}
f_{\textrm{r,peak}}(e) = n_{\textrm{peak}}(e) f_{\textrm{orb}}.
\end{equation}
The value of $n_{\rm peak}$ increases with increasing eccentricity.  To determine which is the peak harmonic for a given value of eccentricity, one can evaluate the entire series of $g(n,e)$ of Eq.~\ref{eq:g_ne}, and find the $n$ that has the maximum value of $g(n,e)$.  However, there is an alternative and faster way to approach this question. Refs.~\cite{Wen:2002km, Hamers:2021eir}, have provided two fitting functions that determine the peak harmonic for a given value of $e$.  We adopt the fitting function from Ref.~\cite{Hamers:2021eir}, as at eccentricities $e \lesssim 0.8$, it corrects for about a  $\sim 10\%$ overestimation error in $n_{\textrm{peak}}$ that Ref.~\cite{Wen:2002km} has compared to the result that comes from using Eq.~\ref{eq:g_ne}.  
Moreover, Ref.~\cite{Hamers:2021eir}, can get the exact  value for $n_{\textrm{peak}}$ for $e$ up to $0.99$ and is accurate to its evaluation of $n_{\textrm{peak}}$ to within $0.01\%$ up to at least $e=0.999$.

The fit for $n_{\rm peak}(e)$ from Ref.~\cite{Hamers:2021eir}, is given by,
\begin{equation}\label{hamers_fit}
n_{\rm peak}(e) = 2 \Biggl(1 + \sum_{k=1}^{4} c_k e^k \Biggr) (1-e^2)^{-3/2},
\end{equation}
with $c_1=-1.01678$, $c_2=5.57372$, $c_3=-4.9271$, and $c_4=1.68506$. 

The characteristic strain amplitude $h_{c,n}$ of the $n$-th harmonic observed at the detector frame $f_{\rm det}(n) = f_{r}(n)/(1+z)$ is 
\cite{Flanagan:1997sx,Barack:2003fp},
\begin{equation}\label{hc,n}
h_{c,n}(f_{\rm det}(n)) =
\frac{(1+z)}{\pi D_L(z)}\sqrt{\frac{2G}{c^3}\frac{dE_n}{df_{r}}}.
\end{equation}
$D_L(z)$ is the luminosity distance to the source binary. In our characteristic strain calculations Eq.~\eqref{hc,n}, the redshift $z$ is evaluated  when the binary reaches ISCO, i.e., $z = z_{\rm ISCO}$. 

The quantity $dE_n/df_{r}$ is the GW energy per bandwidth emitted at the $n$-th harmonic, given by \cite{DOrazio:2018jnv,Huerta:2015pva,Holgado:2020imj,Chen:2016zyo}, 
\begin{equation}
\frac{dE_n}{df_r} =
\frac{G^{2/3}\pi^{2/3}\mathcal{M}^{5/3}}{3
f_r^{1/3}}
\cdot
\left(\frac{2}{n}\right)^{2/3}
\frac{g(n,e)}{F(e)},
\end{equation}
with $\mathcal{M}$ being  the chirp mass of the binary. 

While the total characteristic strain is a sum over all harmonics in quadrature,
\begin{equation}
h_c^2 = \sum_n h_{c,n}^2 \,
\end{equation}
in this work, we approximate the strain as being dominated by the peak harmonic. 

At the early stages of any binary's evolution, the characteristic strain is negligible. 
When testing the sensitivity of any GW observatory to detect PBH binaries, we restrict our analysis to the regime where the GWs from the inspiral enter the detector-sensitive frequency band. 
Depending on the observatory, GW frequencies lie in the range of $10^{-5}\ \rm{Hz} \le f_{\rm det} \le 100\ \rm{Hz}$. 

\subsection{Detectability across various detectors}\label{subsec:SNR}
To study the detectability of  PBH binary inspiral, we follow the methodology provided by  
Ref.~\cite{Flanagan:1997sx}. The sky- and orientation-averaged squared SNR for a detector is given by
\begin{equation}
\left\langle\frac{S^2}{N^2}\right\rangle =F\int_{f_{\textrm{start}}}^{f_{\textrm{end}}} 
\left[\frac{h_c(f_{\textrm{det}})}{h_n(f_{\textrm{det}})}\right]^2 d\ln f_{\textrm{det}}.
\label{eq:snr}
\end{equation}
The noise strain $h_n(f)$ is again $h_n(f)=\sqrt{f\,S_n(f)}$. We take that noise for current and future detectors from the publicly available \texttt{GWplotter}~\cite{Moore:2014lga}. 
The factor $F$ accounts for the binaries' orientation averaging. 
We set the integration limits in Eq.~\eqref{eq:snr}, based on the overlap between the binary's detector-frame frequency range $[f_{\rm bin,min}, f_{\rm bin,max}]$ and the frequency bandwidth of any given detector $[f_{\rm det,min}, f_{\rm det,max}]$, 
\begin{align}
f_{\rm start} &= \max(f_{\rm bin,min},\;f_{\rm det,min}), \\
f_{\rm end}   &= \min(f_{\rm bin,max},\;f_{\rm det,max}).
\end{align}

We consider a PBH binary to be detectable if its $\langle S/N\rangle \geq 8$. The maximum redshift $z_{\rm max}$ at which this value is met determines the detector horizon.
The expected number of detectable inspirals is then
\begin{eqnarray}\label{eq:det}
N_{\rm det}
&=& t_{\rm obs}\int_{0}^{z_{\rm max}} dz\;
\frac{R_{\rm channel}(z)}{1+z}\,\frac{dV_c}{dz} \nonumber\\
&=& 4\pi\,t_{\rm obs}\;
\int_{0}^{z_{\rm max}} dz\,
\frac{c\,\chi^2(z)\,R_{\rm channel}(z)}{(1+z)\,H(z)},
\end{eqnarray}
where $\chi(z)$ is the comoving distance, $H(z)$ is the Hubble parameter, and $R_{\rm channel}(z)$\footnote{For the unperturbed channel this is the total channel rate. For the binary-single interaction channel in using the comoving merger rate, we need to also keep track of the dark matter halo mass range studied.} is the PBH merger rate for any given evolutionary channel as most recently evaluated in~\cite{Aljaf:2025dta}.  
We note that while GW observatories in space will not probe the merger of the stellar mass range black hole binaries, a connection can be made between their observations and those of concurrent ground-based observatories \cite{Sesana:2016ljz}.

\section{Results}\label{sec:Results}
We discuss our results on the possibility of detecting PBH binaries with some eccentricity in the late stages of their inspirals, from each merger channel separately, before combining them. 
\subsection{Unperturbed Binaries}\label{unperturbed_results}
Our initial sample contains $5\times10^6$ binaries at $z=3400$. Of these, $2.5\times10^5$ binaries merge by $z=12$. Thus, we simulate beyond that redshift the remaining $4.75\times10^6$ PBH binaries. 
By $z=0$, $9.08\times10^4$ more binaries reach their ISCO. 

\begin{figure}[ht!]
    \centering
    \includegraphics[width=0.99\linewidth]{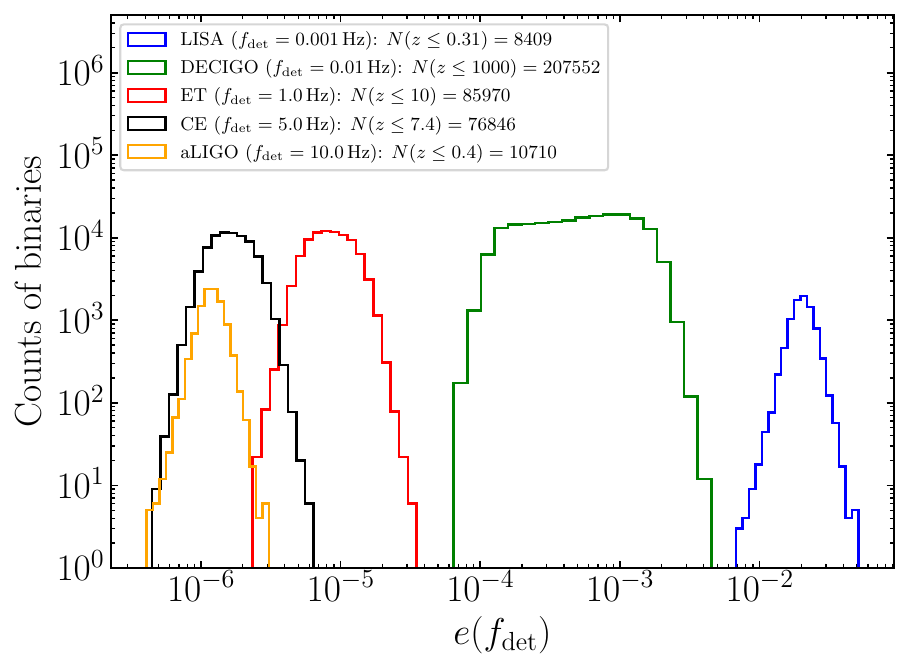}
    \caption{The eccentricity distributions from an initial sample of $5\times10^6$ PBH binaries with $m_1=m_2=30\,M_\odot$, and assuming $f_{\rm PBH}=1$ and $f_{\rm PBH \, binaries}=0.5$, evolving purely through GW emission. These distributions are evaluated at the moment the simulated binaries' emitted GWs reach detector-frame frequencies of $0.001\,\rm{Hz}$, $0.01\,\rm{Hz}$, $1\,\rm{Hz}$, $5\,\rm{Hz}$ and $10\,\rm{Hz}$. We use these frequencies as reference values for LISA, DECIGO, ET, CE and aLIGO, respectively. In each histogram, the binaries that reach the relevant frequency beyond an observatory's maximum horizon redshift are excluded.}
    \label{fig:raw_eccentricity_hist}
\end{figure}

In Fig.~\ref{fig:raw_eccentricity_hist}, out of the $5\times10^6$  initial sample, we show the eccentricity distributions of the unperturbed PBH binaries, evaluated when these binaries reach during their inspirals, detector-frame GW frequencies of $0.001 \,\rm Hz$, $0.01 \,\rm Hz$, $1\,\rm Hz$, $5\,\rm Hz$\, and $10\,\rm Hz$.  
These frequency values are used as references to probe the moment at which the GW emission from PBH binaries' inspirals will reach the  LISA, DECIGO, ET, CE, and aLIGO minimum detector frequencies, respectively. 
These histograms include only binaries merging within each GW observatory's redshift horizon ($z \le z_{\max}$). 
Colors distinguish the GW observatory bands and the legend lists the corresponding binary counts $N(z \le z_{\max})$ meeting this observatory's redshift horizon.
We evaluate the redshift horizons for each observatory by taking a binary of two $30 \, M_{\odot}$ black holes and finding at which redshift such a binary would have a signal-to-noise ratio of 8 (see Section~\ref{subsec:SNR} for more details). Those redshift horizons are given in Table~\ref{tab:detector_specs}.

\begin{figure}[ht!]
    \centering
    \includegraphics[width=0.99\linewidth]{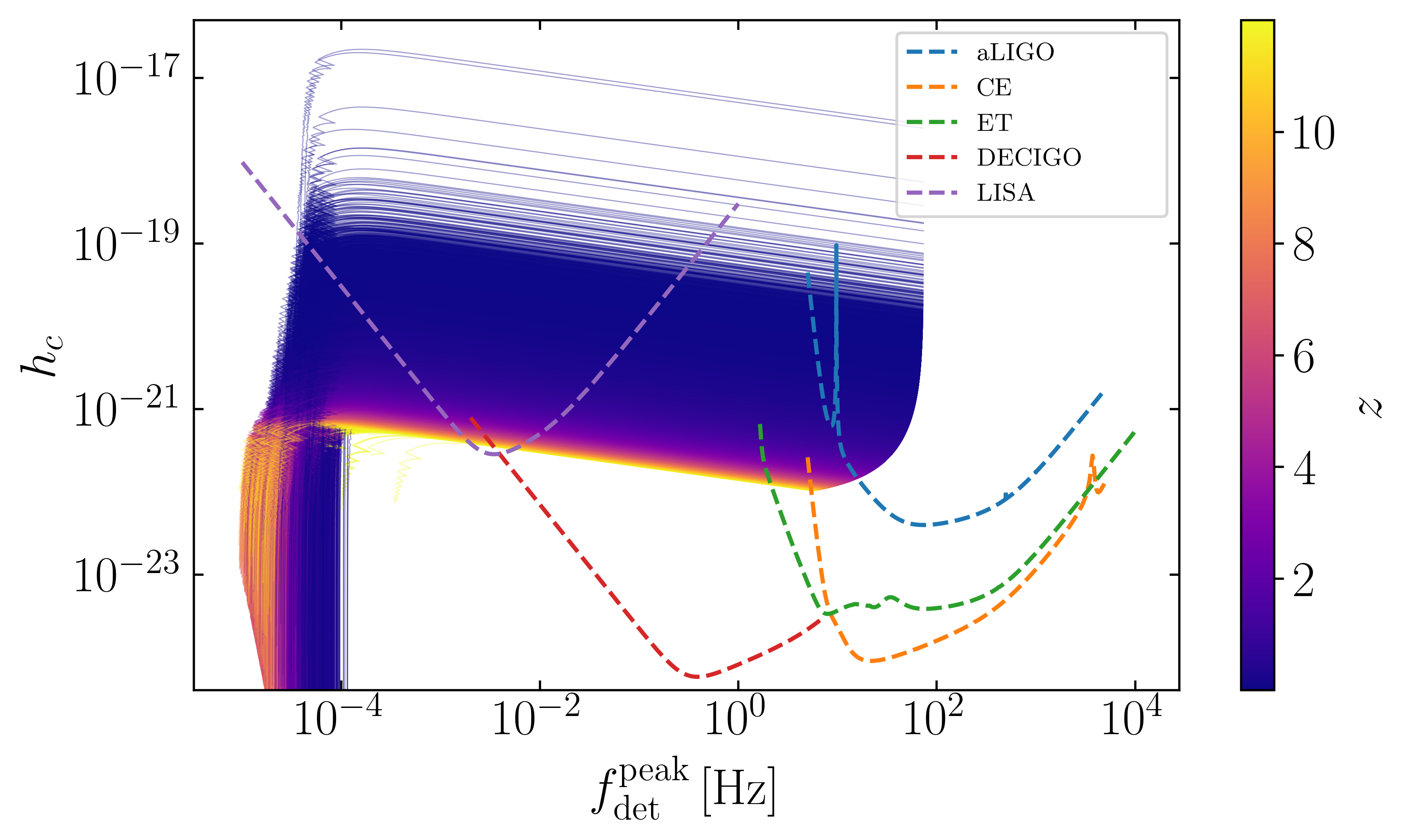}
    \includegraphics[width=0.99\linewidth]{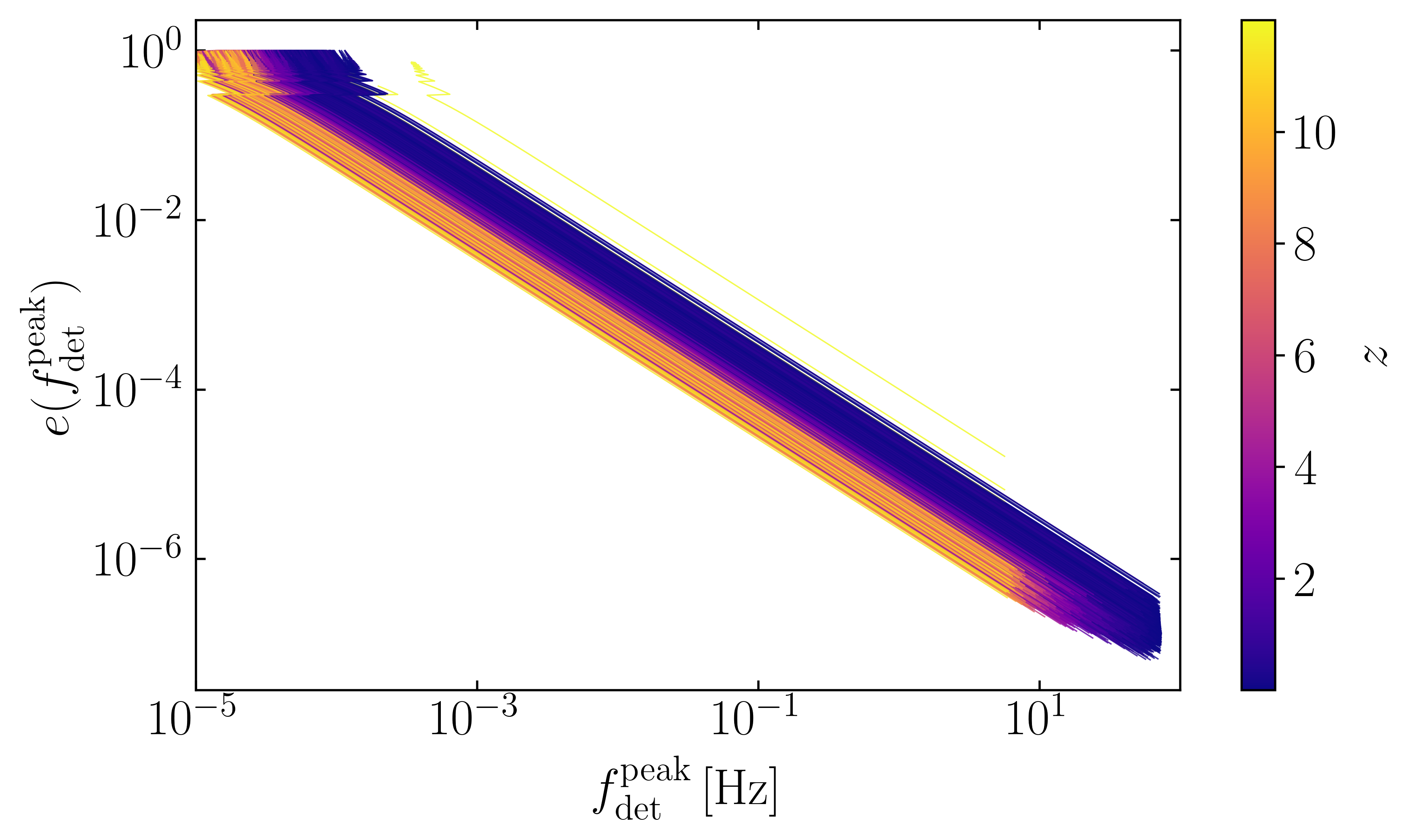}
    \caption{Unperturbed PBH merging binaries ($m_1=m_2=30\,M_\odot$, $f_{\rm PBH}=1$ and $f_{\rm PBH \, binaries}=0.5$), showing the raw, unweighted simulation tracks from $z=12$ to ISCO using GW emission only. Top panel: characteristic strain $h_c$ vs detector-frame frequency $f_{\rm det}^{\rm peak}$, plotted against sensitivity curves from \texttt{GWplotter}~\cite{Moore:2014lga}. Bottom panel: orbital eccentricity $e$ vs $f_{\rm det}^{\rm peak}$. Data points are color-coded by the redshift at which they reach their ISCO. In both panels, binaries that merge above $z=12$ are excluded. }
    \label{fig:unperturbed_binaires}
\end{figure}

Fig.~\ref{fig:raw_eccentricity_hist} shows the effect of GW-driven circularization as the binaries inspiral toward ISCO. At $0.001\,\rm Hz$, corresponding to the GW frequency detectable by LISA, most binaries already have relatively small eccentricities, typically around $10^{-2}$. As the GW frequency increases, the distributions shift toward even lower eccentricities. 
In Fig.~\ref{fig:unperturbed_binaires}, we show how the characteristic strain of the most dominant mode $h_{c,n}$ and eccentricity evolve with GW frequency at detection. We plot the raw simulation tracks without applying any horizon or detectability cuts. As the binaries inspiral, they move from the LISA band through DECIGO and eventually into the frequency range of ET, CE, and aLIGO. Furthermore, GW emission causes the decay of  their eccentricities. 
We track the most dominant mode, which at low frequencies can be seen as the lines having a serrated pattern (as the most dominant GW emission mode changes).
While many binaries still have residual eccentricities of $e\sim10^{-2}$ at the LISA entry frequency and to a smaller extent at the DECIGO frequency, they become nearly circular ($e<10^{-4}$) before reaching the bands of the ground-based observatories.    
The counts for eccentricity distributions in Fig.~\ref{fig:raw_eccentricity_hist}, cannot be compared directly with observations, as they come from a fixed, arbitrary sample of initial $5\times10^6$ binaries. To obtain the observable distributions, we need to account for the merger rate given the sensitivity and detection horizon of each observatory.
For each detector, we calculate its maximum detectable redshift, $z_{\max}$, by requiring $\langle S/N\rangle \geq 8$ using Eq.~\eqref{eq:snr}.  We then evaluate Eq.~\eqref{eq:det} to find the expected number of detectable inspirals, $N_{\rm det}$. 
To compute $N_{\rm det}$, we use the proper unperturbed comoving merger rate $R_{\rm unperturbed}$ from Ref.~\cite{Aljaf:2025dta}, assuming $f_{\rm PBH}=3.4\times 10^{-3}$, $f_{\rm PBH \,binaries}=0.5$, and $m_1=m_2=30,M_\odot$. This combination of PBH parameters is in agreement with the current limits from the LVK Collaboration observations \cite{Bouhaddouti:2026jgc}.
We also test alternative assumptions on the combination of $f_{\rm PBH}$ and $f_{\rm PBH \,binaries}$ that are in agreement with the LVK observations and find the same number of detected events as in Table \ref{tab:detector_specs}.

\begin{figure}[ht!]
    \centering
    \includegraphics[width=0.99\linewidth]{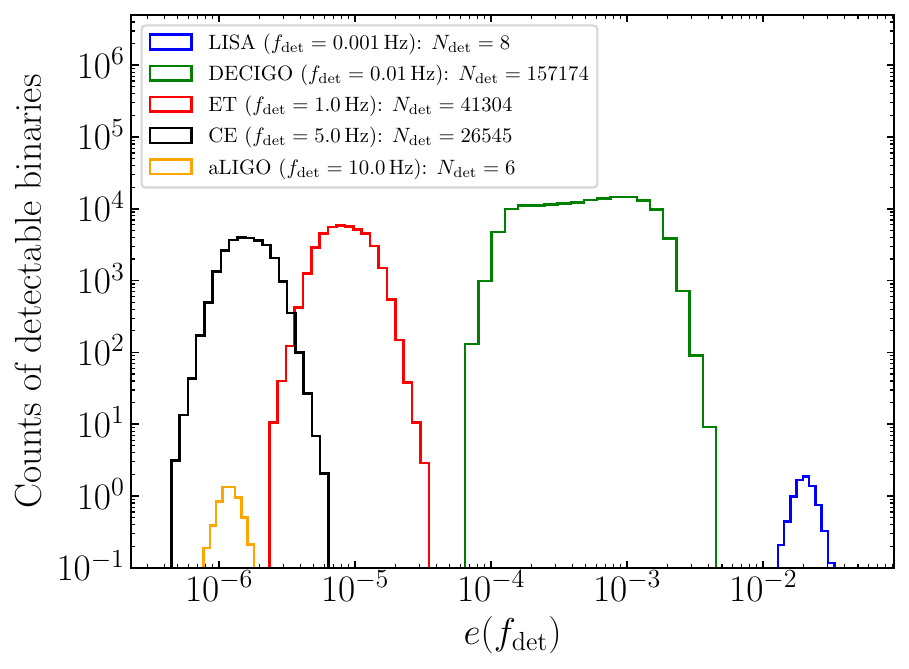}
    \caption{The number of detectable binaries $N_{\rm det}$ with their eccentricity distribution (i.e. Fig.~\ref{fig:raw_eccentricity_hist} properly rescaled). This represents only the unperturbed PBH channel across the different detector frequency bands.}
    \label{fig:weighted_eccentricity_hist}
\end{figure}

The resulting values of $z_{\max}$ and $N_{\rm det}$ are listed in Table~\ref{tab:detector_specs}. LISA and aLIGO are sensitive mainly to low redshift, whereas ET and CE reach a redshift of 10 and 7.4, respectively. DECIGO can reach even higher redshifts than $z_{\max}=1000$, but to be conservative, we cap our calculations to that value. 
\begin{table}[h!]
\centering
\begin{tabular}{lccc}
\hline
Detector & $z_{\max}$ & $t_{\rm obs}$ (yr) & $N_{\rm det}$ \\
\hline
LISA   & 0.31 & 5  & 8 \\
DECIGO & 1000  & 3  & 157174 \\
ET     & 10   & 10 & 41304 \\
CE     & 7.4    & 10 & 26545 \\
aLIGO  & 0.4  & 2  & 6 \\
\hline
\end{tabular}
\caption{Detector horizon redshifts $z_{\max}$ computed via Eq.~\eqref{eq:snr}, along with assumed observation times $t_{\rm obs}$, and the resulting expected number of detectable PBH binary inspirals $N_{\rm det}$ from Eq.~\eqref{eq:det}.} 
\label{tab:detector_specs}
\end{table}
The final rescaled  distributions for unperturbed binaries are shown in Fig.~\ref{fig:weighted_eccentricity_hist}. While LISA and aLIGO will observe a similar number of unperturbed PBH binaries, the aLIGO binaries will be fully circularized. The binaries by LISA instead will have some minor remaining eccentricity at the $O(10^{-2})$ level. DECIGO, ET, and CE can observe many more unperturbed binaries. 
However, none of these binaries will have any substantive remaining eccentricity.

\subsection{Binary single interactions}
As we showed, unperturbed PBH binaries almost completely circularize by the time their emitted GWs reach ground-based-detector frequencies. However, PBH binaries residing in dark matter halos can undergo binary-single interactions with neighboring PBHs that change their orbital evolution, increasing their orbital eccentricity. 
We evolve a sample of $4.75 \times 10^6$ PBH binaries inside each radial shell of a given halo from $z = 12$ down to $z = 0$. These simulations are carried out across five halo masses: $1.68 \times 10^4\,M_\odot$, $1.60 \times 10^5\,M_\odot$, $1.53 \times 10^6\,M_\odot$, $1.46 \times 10^7\,M_\odot$, and $1.39 \times 10^8\,M_\odot$, each serving as a representative dark matter halo for all halos within its respective mass bin.

Based on our earlier work in Ref.~\cite{Aljaf:2025dta}, we know that for halos in the range of $[10^9,10^{15}]\,M_\odot$,  we can treat them as effectively unperturbed binaries that follow the same eccentricity distributions for the given frequencies. 
While mergers still happen inside those more massive halos, the binary-single interactions in those environments are rare and have a negligible effect on the orbital evolution of the PBH binaries. 

For a given halo we build histograms of eccentricity distribution in three steps: (i) we evolve binaries within each mass shell of a given halo mass and record the raw eccentricity distribution of those that merge, excluding the binaries that merged beyond the detector's horizon redshift $z_{\rm max}$, (ii) we combine the shell's distributions into one distribution for the whole halo, weighted by each shell's mass-fraction contribution and (iii) we rescale that distribution using the number of detectable binaries $N_{\rm det}$, from the comoving merger rate of the halo's mass bin (relying on the rates from Ref.~\cite{Aljaf:2025dta}). 

\begin{figure}[ht!]
    \centering
    \includegraphics[width=0.935\linewidth]{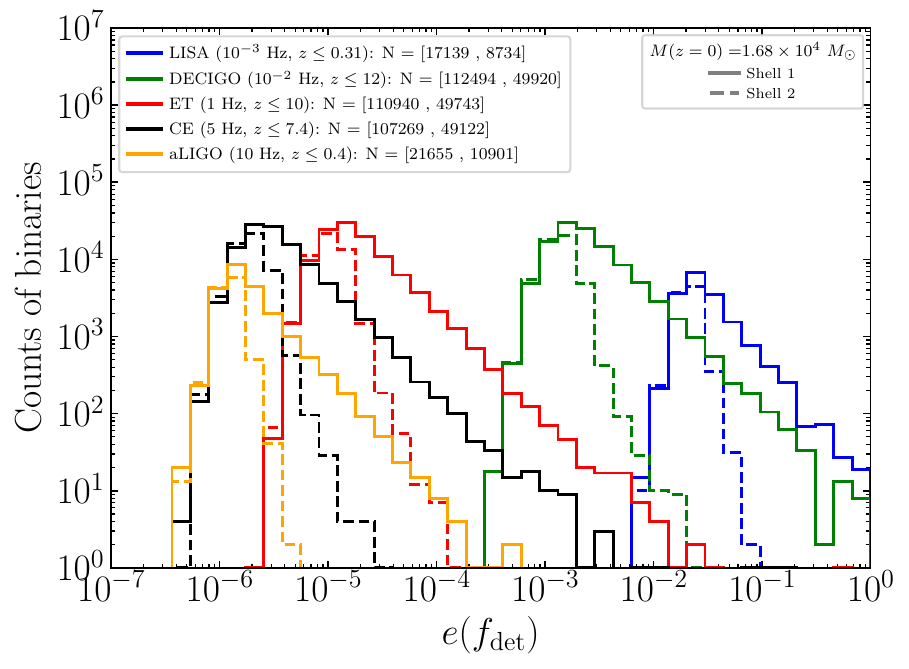}
    \includegraphics[width=0.935\linewidth]{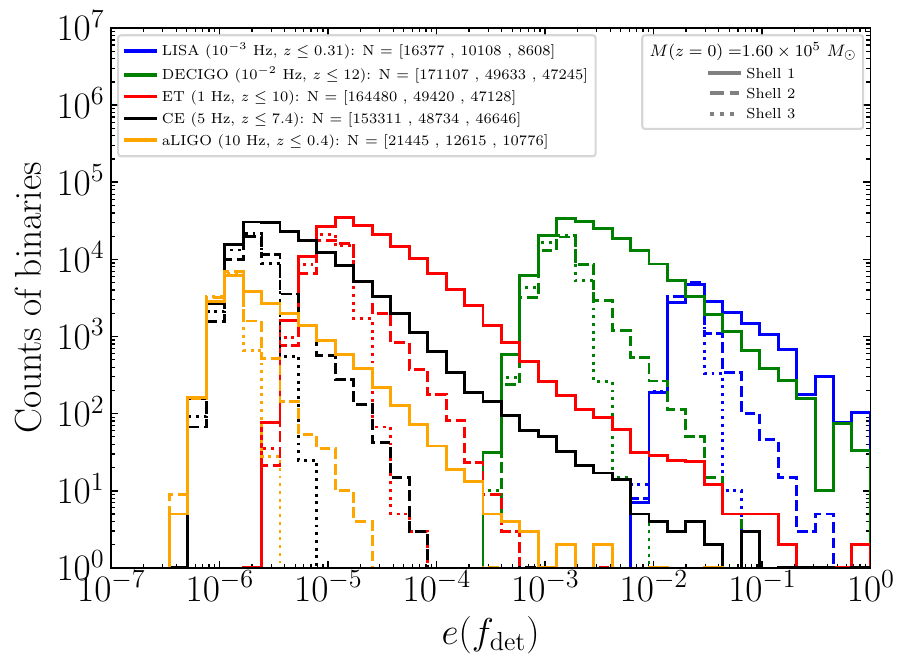}
    \includegraphics[width=0.935\linewidth]{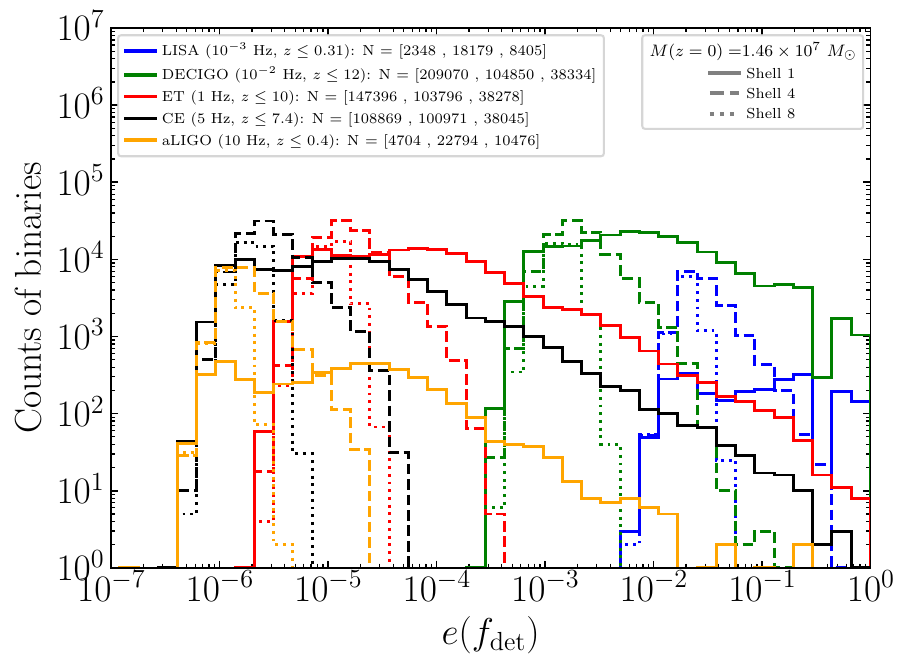}
    \caption{Eccentricity distributions of PBH binaries from the binary-single interaction channel, shown separately for each shell of three representative halos, with present-day masses $M(z=0) = 1.68\times10^4\,M_\odot$ (top: two shells), $1.60\times10^5\,M_\odot$ (middle: three shells), and $1.46\times10^7\,M_\odot$ (bottom: ten shells, of which only shells ``1'', ``4'', and ``8'' are shown). Colors are the same as in Figs.~\ref{fig:raw_eccentricity_hist} and~\ref{fig:weighted_eccentricity_hist}, distinguishing the detector band at which $e$ is evaluated. The line style distinguishes the shell, with solid, dashed, and dotted lines corresponding to increasingly outer shells (solid always being for shell ``1'', the innermost and densest one). Only binaries merging within a given observatory's redshift horizon ($z\leq z_{\max}$) are included. The legend lists the raw counts $N$ out of $4.75 \times 10^6$ simulated binaries in each shell for each observatory.}
    \label{fig:Mh_shells_eccentricity_hist}
\end{figure}

Fig.~\ref{fig:Mh_shells_eccentricity_hist}, we present the results for the eccentricity distributions of PBH binaries vs detector-frame frequency, for three selected halos, with present-day masses of $1.68\times 10^4 \, M_{\odot}$, $1.60\times 10^5 \, M_{\odot}$, and $1.46\times 10^7 \, M_{\odot}$. 

The halo of mass $1.68\times 10^4\,M_\odot$ is divided into 2 shells (top panel),  $1.60\times 10^5\,M_\odot$ into 3 shells (middle panel), and $1.46\times 10^7\,M_\odot$ into 10 shells, out of which only shells 1, 4, and 8 are shown in the histograms (bottom panel). 

\begin{figure}[ht!]
    \centering
    \includegraphics[width=0.96\linewidth]{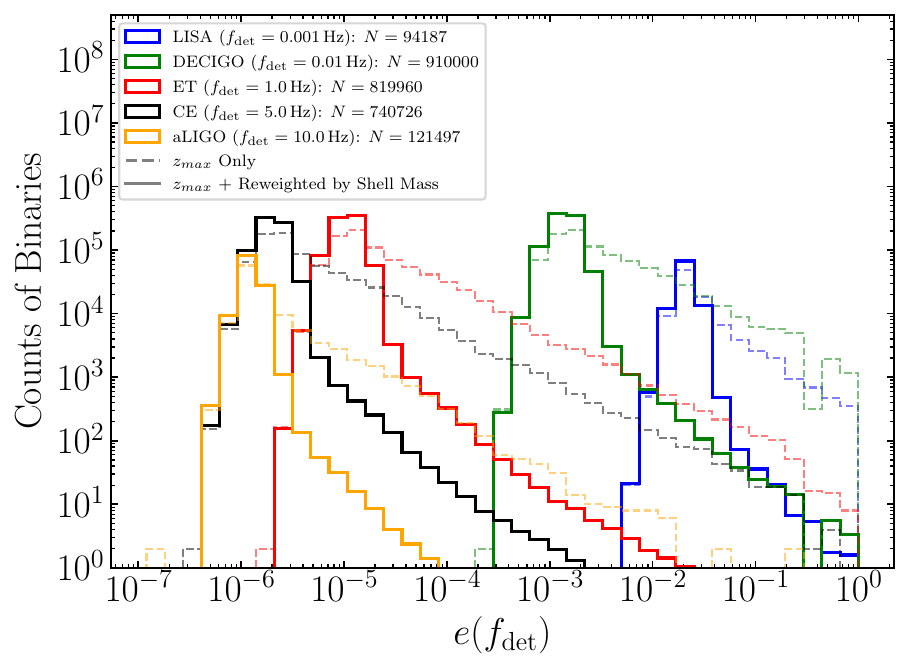}
    \includegraphics[width=0.99\linewidth]{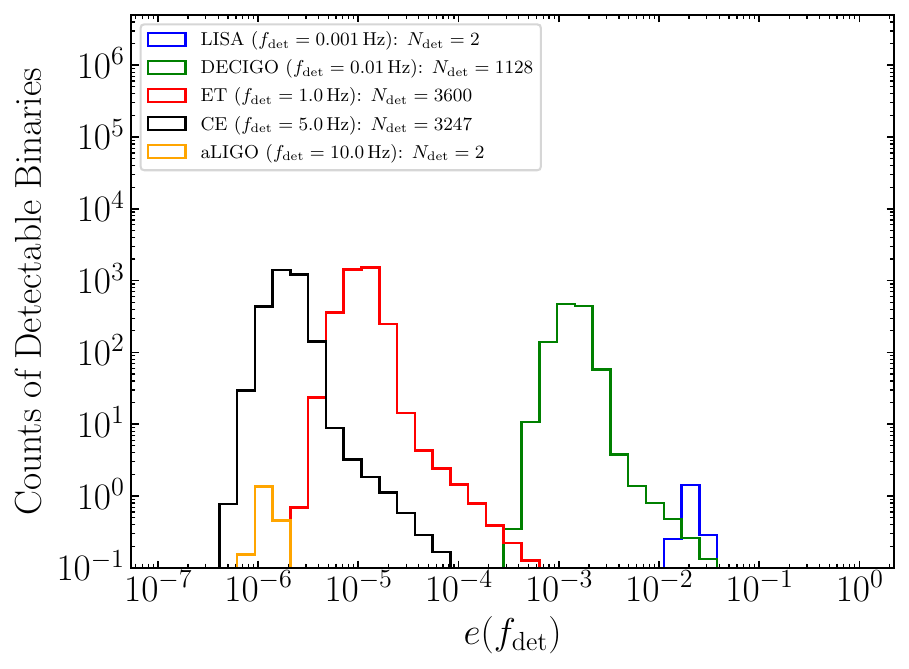}
   \caption{The eccentricity distribution from the binary-single interaction channel for the $M(z=0)=1.46\times10^7\,M_\odot$ halo, combining all shells and applying the detector horizon redshift cut ($z\le z_{\max}$) and shell mass weighting. Top: comparison between the unweighted sum over all shells and the distribution weighted by each shell's mass fraction using Eq.~\eqref{eq:weight}. Bottom: the shell mass-weighted distribution is rescaled to the expected number of detectable binaries $N_{\rm det}$ using  Eq.~\eqref{eq:det}.}
    \label{fig:ecc_weighted_comparison}
\end{figure}

We use the same colors as in Figs.~\ref{fig:raw_eccentricity_hist} and~\ref{fig:weighted_eccentricity_hist}, to distinguish the detector frequency at which $e$ is evaluated. 
Solid lines correspond to the inner shell (``1''), and each subsequent style corresponds to a shell at a larger radius. 
The legend also lists the counts $N$ of the binaries reaching ISCO within $z_{\max}$, given separately for each shell and each detector.
For the smallest halo we consider, $1.68\times10^4\,M_\odot$, the relative fraction of counts between the innermost shell (shell 1) and the outermost shell (shell 2), $N_1/N_2$, stays similar for different observatories.  
For instance, $N_1/N_2= 1.96$ at LISA, $2.25 $ at DECIGO, $2.23$ at ET, $2.18 $ at CE, and $1.98$ at aLIGO.  
The fact that this fraction is bigger than 1 is due to higher efficiency of the binary-single interactions  inside shell ``1'' compared to those residing inside shell ``2'', which results from  higher densities and lower dispersion velocities. 
These interactions make it twice as likely for a PBH binary inside shell ``1'' to merge compared to a PBH binary inside shell ``2''. 
We note again that the presented numbers in Fig.~\ref{fig:Mh_shells_eccentricity_hist}, have not yet been properly weighted by the relative dark matter mass included in each mass shell
(which we account for in Fig.~\ref{fig:ecc_weighted_comparison}). 

The differences between the inner shell and the outer shells become more prominent for more massive halos. 
For a $1.6\times10^5\,M_\odot$ halo mass, the relative fraction of counts between its innermost shell and the outermost shell (shell ``1''/shell ``3''), $N_1/N_3$, changes to $= 1.90$ at LISA, $3.62 $ at DECIGO, $3.49 $ at ET, $3.29 $ at CE, and $1.99$ at aLIGO.  This increase is due to the fact that the efficiency gradient of binary-single interactions between the innermost shell and the outermost shell (shell ``1'' vs shell ``3'') in this halo mass is much steeper than the gradient between shell ``1'' and shell ``2'' in the less massive halo of $1.68\times10^4\,M_\odot$.

For the $1.46\times10^7\,M_\odot$ halo, the relative fraction of counts between its innermost shell (shell ``1'') and  one of the outer shells (shell ``8''), $N_1/N_8$, is $0.28$ at LISA, $5.45 $ at DECIGO, $3.85 $ at ET, $2.86 $ at CE, and $0.45$ at aLIGO.  
The fact that $N_1/N_8$ is much larger than one for DECIGO, ET, and CE, but less than one, is due to the extreme properties of shell ``1'' compared to shell ``8''. In the innermost shell, the chance that a PBH binary will merge is quite high, but at the same time as the evolution is faster, many of these mergers happen at high enough redshifts that they will be outside the reach of LISA and aLIGO (i.e., $z_{\max} = 0.31$ and 0.4, respectively).
The efficiency of binary single-interactions determines the number of merged binaries and their redshift, but whether they would be detectable depends on the observatory's horizon redshift. This also determines the shape of the histograms from each shell shown in Fig.~\ref{fig:Mh_shells_eccentricity_hist}.

To combine the eccentricity distributions of binaries of individual mass shells of a given halo mass into one eccentricity distribution that represents the whole halo mass, we need to weight the histograms from each shell by their relative shell mass fraction. We define the weight belonging to the $j$-th shell of a halo of mass $M$ as,
\begin{equation}\label{eq:weight}
w_j = \frac{f_j N(z < z_{\mathrm{max}})}{\sum_{j} N_j(z < z_{\mathrm{max}}) f_j},
\end{equation}
where $N_j$ is the number of binaries that merged in our simulation in the $j$-th shell and $N$ is the total number of merged binaries combined across all shells. 
The parameters $f_j$ are the mass fractions from each shell (indexed by ``j'') relative to the total mass of the halo $M$. They are defined as, 
\begin{eqnarray}
f_j=\frac{M_{j}(z)}{M(z)}.
\end{eqnarray}
In Fig.~\ref{fig:ecc_weighted_comparison} (top panel), we present the effect of this weighting for the $1.46\times10^7\,M_\odot$ halo. In our simulations, most of the halo mass is in the outer shells. Thus, the contribution of inner shells is comparatively small. This suppresses the contribution of inner shells and shifts the combined eccentricity distribution toward lower $e$ values, characteristic of the binaries merging in the outer regions.
In Appendix~\ref{app:other_halo_masses}, 
we provide the equivalent results of the PBH binaries from the binary-single channel eccentricity distributions from the other simulated halo masses.  
To evaluate the eccentricity distribution relative to the expected number of detectable binaries from halos within a mass range, we rely on Eq.~\eqref{eq:det}. For that, we need the comoving merger rate contributed by the halos that fall withihn its corresponding mass bin $[M_k, M_{k+1})$, which we take from Ref.~\cite{Aljaf:2025dta}. 
The bottom panel of Fig.~\ref{fig:ecc_weighted_comparison} shows the resulting distribution for the $1.46\times10^7\,M_\odot$ halo after being rescaled with $N_{\rm det}$.
In  Appendix~\ref{app:other_halo_masses}, we provide the properly merger-rate weighted eccentricity distributions of PBH binaries from the binary-single channel, for the other simulated halo masses.  

\begin{figure}[ht!]
    \centering
    \includegraphics[width=0.99\linewidth]{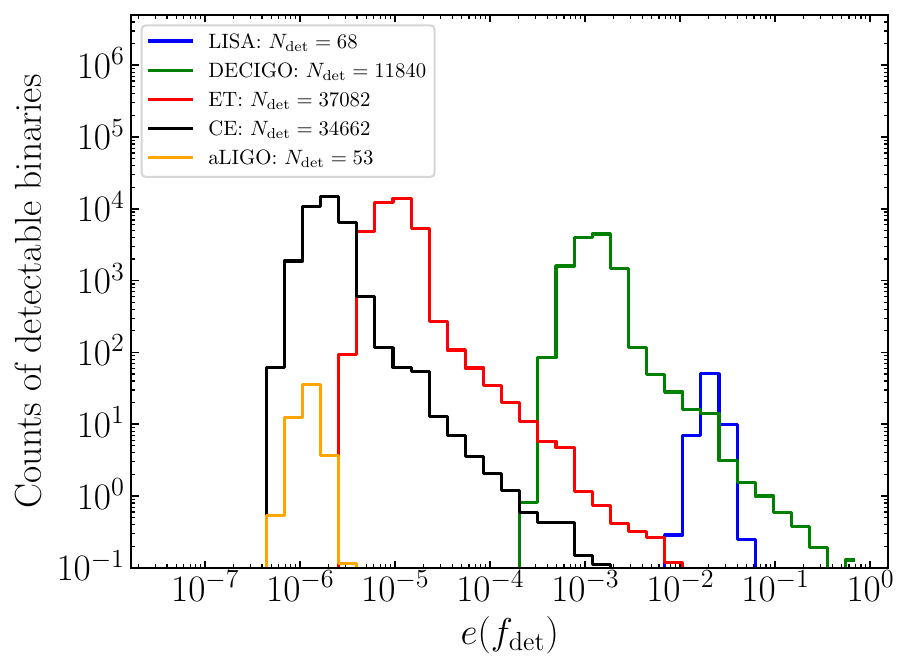}
    \caption{Eccentricity distributions for the binary-single
    interaction PBH channel across different detector frequency bands,
    combining all simulated halo masses ($10^4$--$10^{15}\,M_\odot$),
    weighted by shell mass fraction and scaled to $N_{\rm det}$.}
    \label{fig:Perturbed_binary_single_ecc_hist}
\end{figure}

For halos in the mass bin of $[10^9  ,10^{15} ] \, M_\odot $, binary-single interactions are negligible, so we instead adopt the unperturbed eccentricity distribution (Sec.~\ref{unperturbed_results}) directly scaled via Eq.~\eqref{eq:det}. Combining the scaled histograms from all five representative halo masses together with the rescaled unperturbed contribution from the $10^9$–$10^{15}\,M_\odot$ range yields the total binary-single-interaction eccentricity distribution shown in Fig.~\ref{fig:Perturbed_binary_single_ecc_hist}.

\subsection{GW captures}

PBH binaries formed by direct GW captures may lead to very short-lived binaries with extreme eccentricities even at their final stages \cite{Cholis:2016kqi}. However, given the current limits on the $f_{\textrm{PBH}}$, which require that $f_{\textrm{PBH}} \leq 2 \times 10^{-2}$ \cite{Bouhaddouti:2026jgc}, we find that none of the current or future observatories will see PBH binaries with a high eccentricity as a result of a direct capture with subsequent merger. 
For instance, for aLIGO and LISA we get $O(10^{-3})$ events with $e>0.1$ from the direct capture channel. 
For ET and CE, we expect $O(10^{-2})$ such events with $e>0.1$. 
Even with DECIGO, we expect less than one event with $e>0.1$ from direct captures. 

\begin{figure}[ht!]
    \centering
    \includegraphics[width=0.99\linewidth]{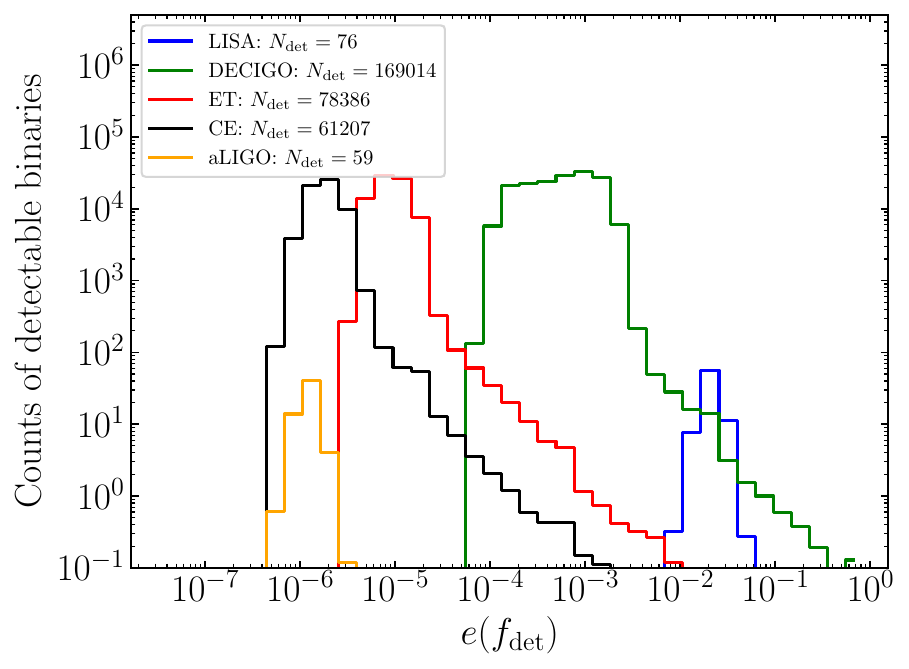}
    \caption{The eccentricity distribution of PBH binaries from all channels. While the binaries detected by ground-based observatories will be effectively circularized, the observatories in space will be able to probe binaries with eccentricities $e > 10^{-2}$ and, in a few rare cases, even $e>0.1$.}
    \label{fig:PBH_total_ecc+per+BS}
\end{figure}

In Fig. \ref{fig:PBH_total_ecc+per+BS}, we combine all PBH merger channels.
The final eccentricity distribution of PBH binaries is effectively the combination of the unperturbed channel and the binary-single interaction channel, with their individual histograms provided in Figs.~\ref{fig:weighted_eccentricity_hist} and~\ref{fig:Perturbed_binary_single_ecc_hist} respectively.

\section{Conclusions}\label{sec:Conclusions}
Stellar mass range PBHs may still contribute a small fraction of the detected black hole binaries by aLIGO.
PBH binaries have large orbital eccentricities at their formation and can retain for a large fraction of their lifetime some of that eccentricity (see e.g. \cite{Sasaki:2016jop, Cholis:2016kqi}).  
In this paper, we explore the prospects that current and future GW observatories have in detecting PBH binaries with some orbital eccentricity even by the time their emitted GWs enter the relevant detector frequency band.
We study the prospects of detections of such binaries by aLIGO, ET, CE, DECIGO, and LISA. 

As the GW observatories probe the stellar-mass range black hole binaries at the late stages of their inspirals, we focus on PBHs that will merge, relying on updated calculations of their merger rates \cite{Aljaf:2025dta}. 
Our calculations study the PBH binaries that formed at effectively the same time the PBHs formed themselves, and then subsequently evolved in isolation unperturbed from interactions with other objects. We also simulate PBH binaries that formed early on but at some point in their history became part of a dark matter halo and gradually ended up in a dense environment where they stochastically interact with other close-by PBHs. 
Binaries of PBHs that fall in the second class may undergo several binary-single interactions that can subsequently harden the binaries and, crucially for this work, increase their orbital eccentricity (see discussion in Section~\ref{sec:PBH orbital evolution}). 
Following our earlier work in Refs.~\cite{Aljaf:2024fru} and~\cite{Aljaf:2025dta}, we simulate PBH binaries belonging to different dark matter halo masses and in different locations within these dark matter halos to properly account for the variety of environmental conditions in which PBH binaries exist. 
We also include the case where PBH binaries form inside dark matter halos from direct captures due to GW emission in close encounters among individual PBHs. 

Since we are interested in binaries that have significant orbital eccentricities while being in their late inspiral stages, in studying their emitted GWs we keep track of the power emitted in different modes, paying special attention to the GW mode with the highest power, i.e., the peak harmonic (see Section~\ref{sec:Characteristic strain}). 
We focus on the PBH binaries that in order to be detected by an observatory they would give a GW signal significantly above the expected noise curves (discussed in Section~\ref{subsec:SNR}). 

Given the current limits on stellar mass PBHs, the aLIGO at its final design sensitivity, binary black hole sample may include several PBH binaries. The binary black hole samples from the ET and the CE observatories may include as many as several tenths of thousand PBH binaries. However, all those PBH binaries will be effectively circularized by the time they emit in such frequencies as to be detected by any of these three ground-based observatories, 
Thus, we will not be able to use the orbital eccentricity as a discriminant from more conventional astrophysical formation mechanisms of binary black holes (see Figs.~\ref{fig:weighted_eccentricity_hist},~\ref{fig:Perturbed_binary_single_ecc_hist} and~\ref{fig:PBH_total_ecc+per+BS}). 
However, with DECIGO and even more so with LISA, we expect to have binaries with orbital eccentricities $e > 10^{-2}$ 
and in a few rare cases even $e>0.1$ by the time their emitted GWs enter the relevant bands (see Fig.~\ref{fig:PBH_total_ecc+per+BS}).

While rare even for the case of PBH binaries, identifying a few binaries with orbital eccentricities of $e \simeq 0.1$ may present a very intriguing result, as binaries from conventional stellar mass black holes are expected to be almost entirely circularized. If however DECIGO and LISA can exclude black hole binaries with $e>0.01$, then the limits on the PBH abundance can be improved by an order of magnitude.

\acknowledgments
We thank David Garfinkle for valuable discussions during the progress of this project. MA and IC have been supported by the National Science Foundation, under grant PHY-2207912. 

\appendix
\section{Shell mass weighting and $N_{\rm det}$ rescaling for other halo masses.}
\label{app:other_halo_masses}
In this appendix, we provide the eccentricity distributions from the binary-single interaction channel for the remaining dark matter halo masses not shown in the main text. We follow the same procedure discussed and detailed in Sec.~\ref{sec:Results} for the $1.46\times10^7\,M_\odot$ halo.  

\begin{figure*}[h!]  
    \centering
    \includegraphics[width=0.48\linewidth]{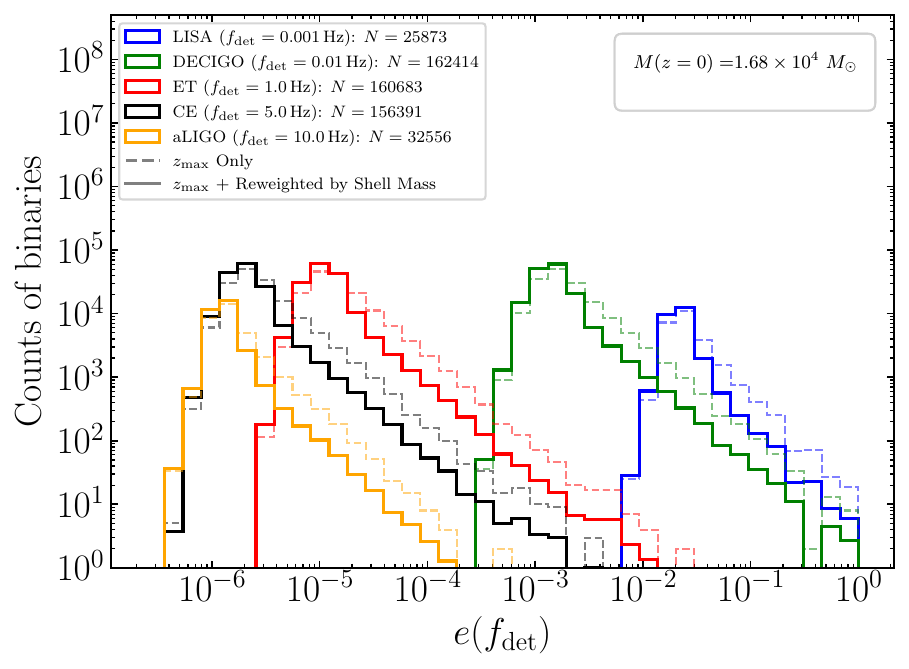}\hfill
    \includegraphics[width=0.48\linewidth]{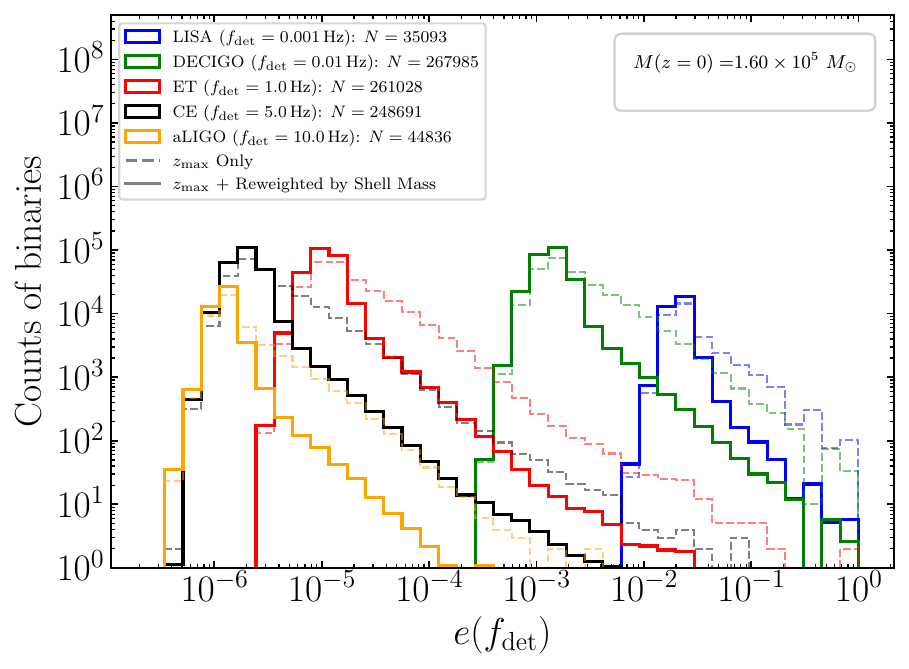}
    \vspace{0.5em}
    \includegraphics[width=0.48\linewidth]{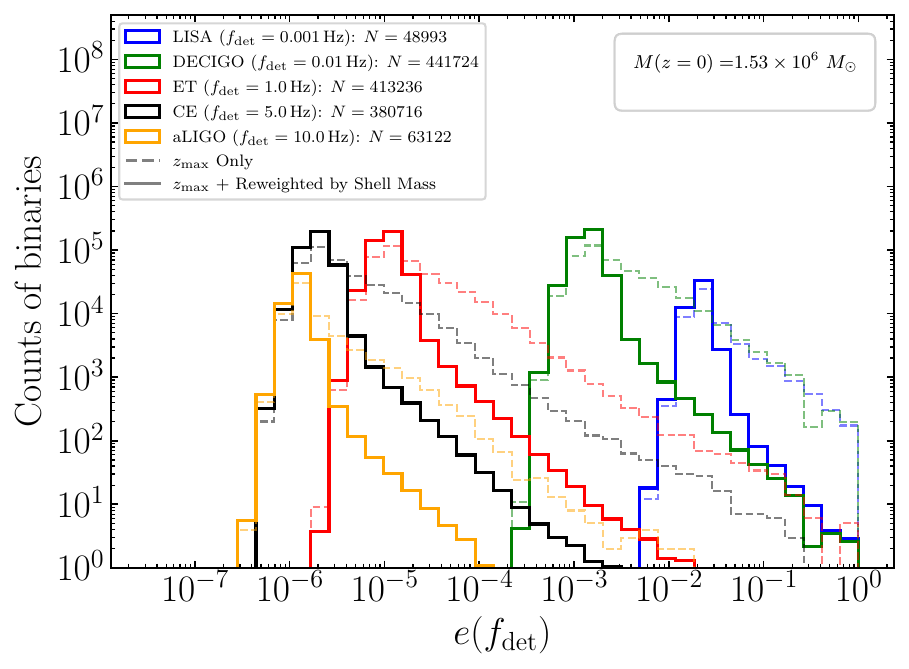}\hfill
    \includegraphics[width=0.48\linewidth]{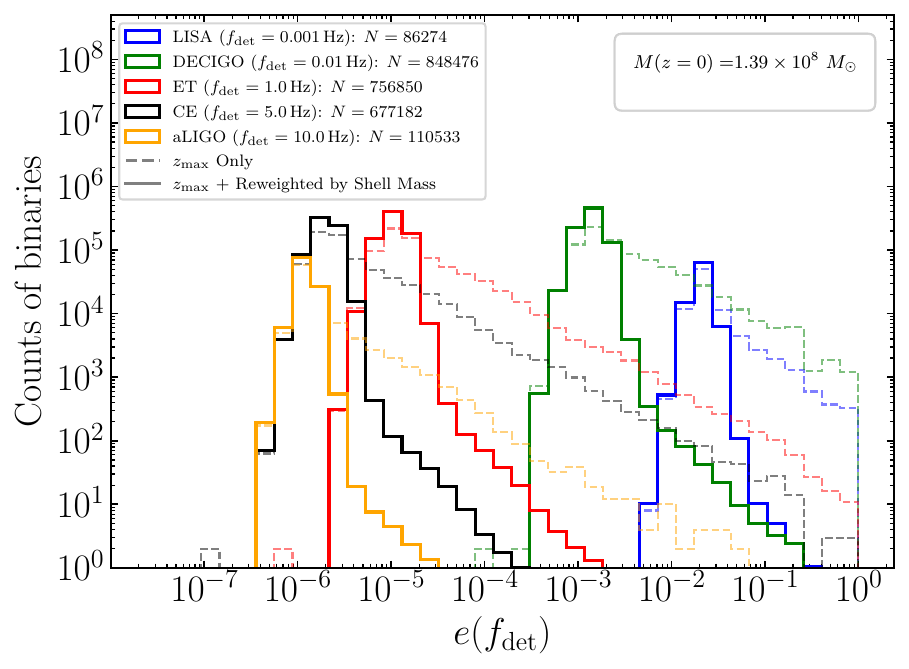}
    \caption{The eccentricity distribution for binary-single interaction channel for halo masses:$M(z=0) = 1.68\times10^4\,M_\odot$ (top left), $1.60\times10^5\,M_\odot$ (top right), $1.53\times10^6\,M_\odot$ (bottom left), and $1.39\times10^8\,M_\odot$ (bottom right). In each panel, the dashed lines are the unweighted sum over all the shells, while the solid lines show the distribution after the re-weighting by each shell's mass fraction according to Eq.~\eqref{eq:weight}. These panels are analogous  to the top panel of Fig.~\eqref{fig:ecc_weighted_comparison} for a $1.46\times10^7\,M_\odot$ halo.}
    \label{fig:ecc_comparison_all}
\end{figure*}

\begin{figure*}[h!]
    \centering
    \includegraphics[width=0.48\linewidth]{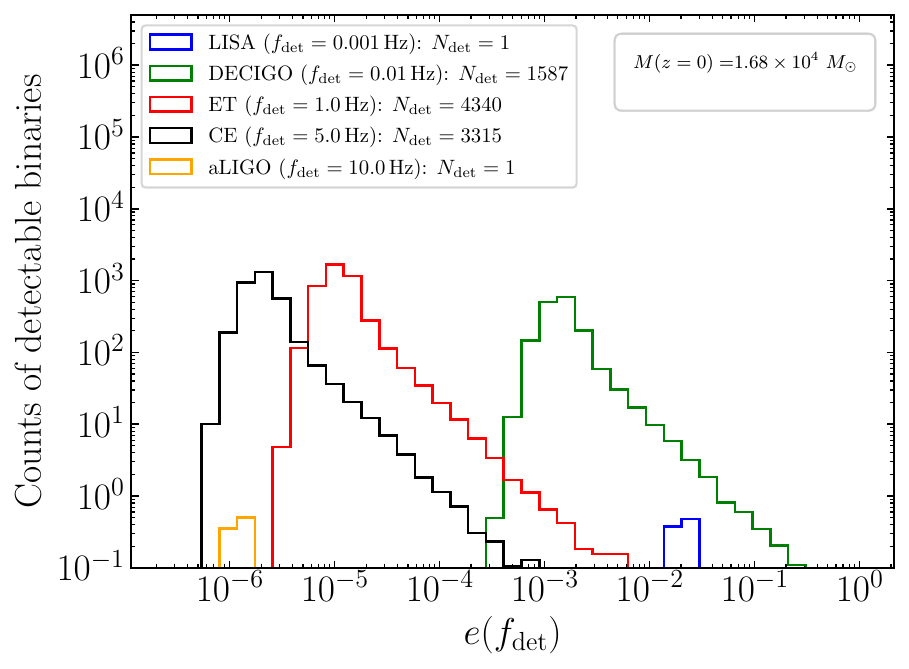}\hfill
    \includegraphics[width=0.48\linewidth]{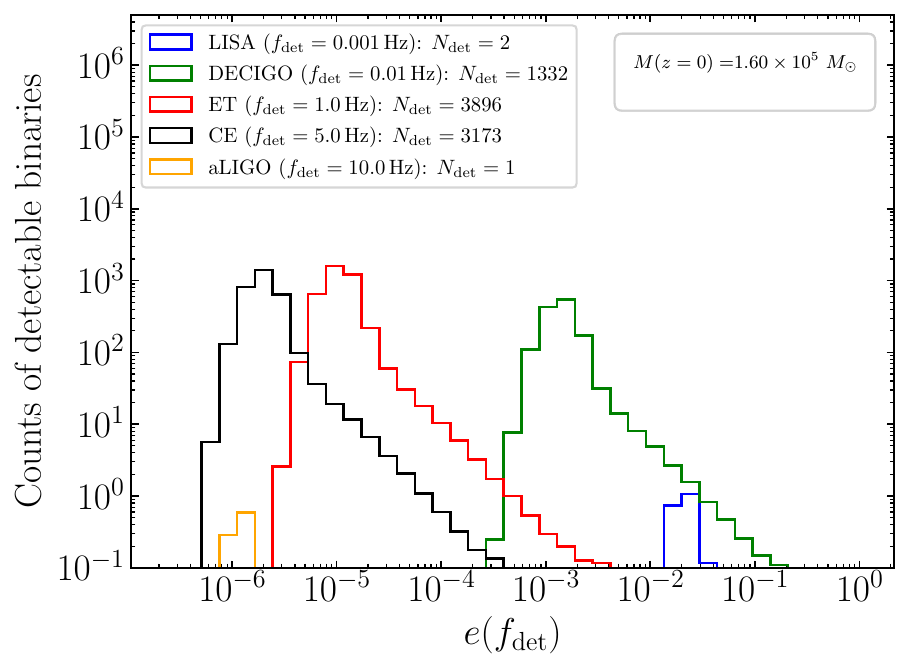}
    \vspace{0.5em}
    \includegraphics[width=0.48\linewidth]{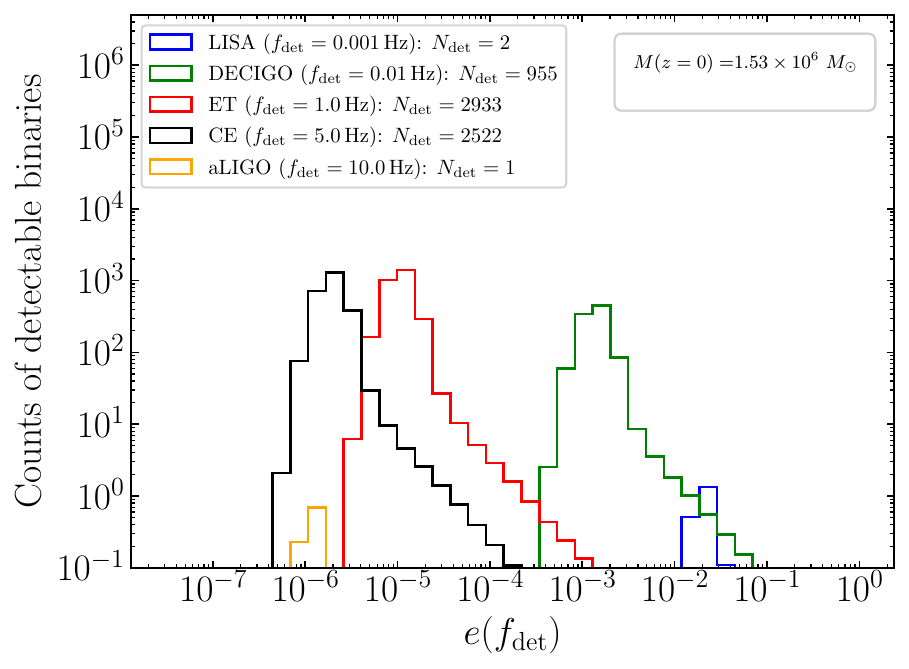}\hfill
    \includegraphics[width=0.48\linewidth]{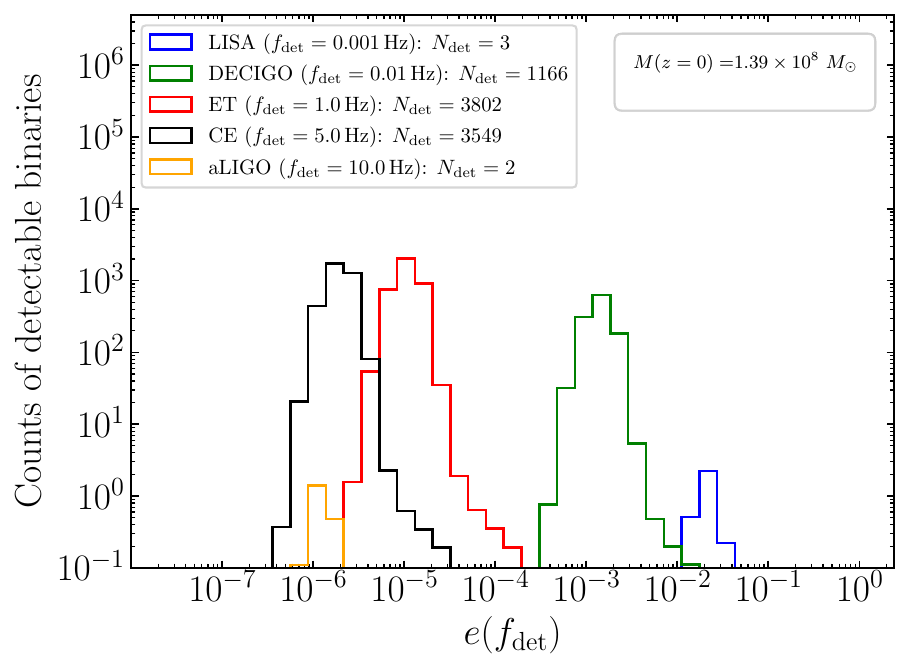}
    \caption{The shell mass-weighted eccentricity distributions of  Fig.~\ref{fig:ecc_comparison_all}, rescaled to the number of detectable PBH binaries $N_{\rm det}$ using  Eqs.~\eqref{eq:det}. This is analogous to bottom panel of Fig.~\ref{fig:ecc_weighted_comparison}, for $1.46\times10^7\,M_\odot$ halo.}
    \label{fig:ecc_det_scaled_halos}
\end{figure*}

In Fig.~\ref{fig:ecc_comparison_all} we show, for each of the four halo masses $M(z=0) = 1.68\times10^4$, $1.60\times10^5$, $1.53\times10^6$, and $1.39\times10^8\,M_\odot$, the effect of combining the eccentricity distribution of binaries in individual shells of a given halo into a single eccentricity distribution: dashed lines represents the $z_{\rm max}$ cut, sum over all shells of a given halo, while solid lines show the same distribution after  being weighted by each shell's mass fraction using Eq.~\eqref{eq:weight}. 
In Fig.~\ref{fig:ecc_det_scaled_halos}, we show these same $z_{\max}$ cut, shell-mass-weighted distributions (solid lines in Fig.~\ref{fig:ecc_comparison_all})), after rescaling to the the number of the detectable binaries $N_{\rm det}$ via Eq.~\eqref{eq:det}, using for each halo's mass bin the comoving merger rate taken from Ref.~\cite{Aljaf:2025dta}.

\begin{figure}[h!]
    \centering
    \includegraphics[width=0.9\linewidth]{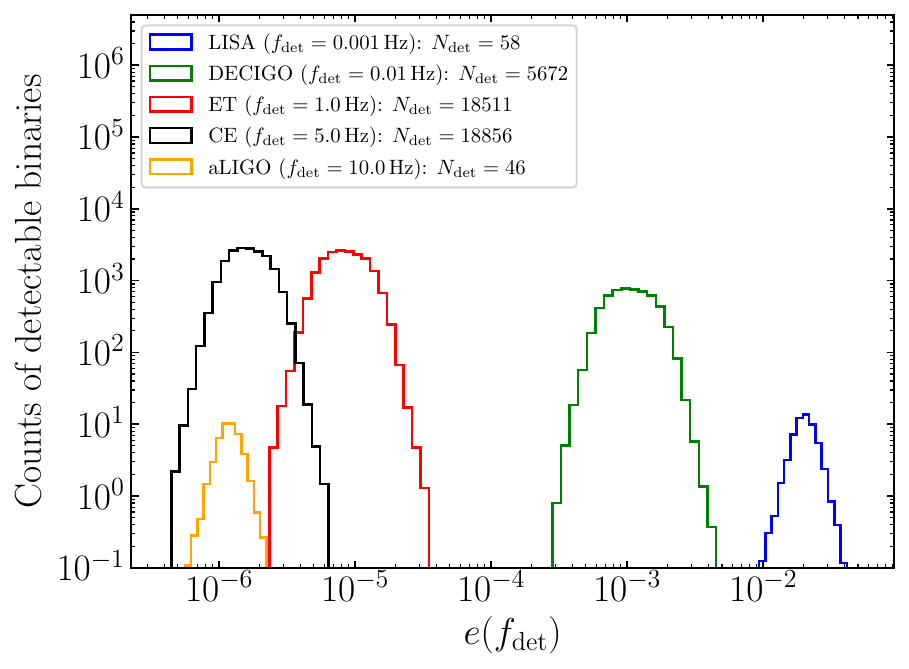}
    \caption{The eccentricity distributions from the halos with the present mass range of $10^9- 10^{15}\,M_\odot$. As binary-single interactions are very rare in halos of this mass range, we adopt the unperturbed distribution of Sec.~\ref{unperturbed_results}, Fig.\ref{fig:raw_eccentricity_hist}, and rescale it with the number of detectable binaries using the comoving merger rate for these halo masses from Ref.~\cite{Aljaf:2025dta} using Eq.\eqref{eq:det}.}
    \label{fig:10to9_to_10to15_halos}
\end{figure}

Finally, in Fig.~\ref{fig:10to9_to_10to15_halos}, we show the analogous $N_{\rm det}$-rescaled, $z\le z_{\max}$ cut distribution for halos in the $10^9$--$10^{15}\,M_\odot$ mass bin, obtained using the comoving merger rate for that mass range from the same reference, for which binary-single interactions are very rare and the unperturbed eccentricity distribution of Sec.~\ref{unperturbed_results} is adopted directly.

\appendix
\bibliography{REF.bib}
\end{document}